\documentclass[preprint, superscriptaddress,preprintnumbers,amsmath,amssymb,prd]{revtex4}
\usepackage{graphicx}

\begin{document}

\thispagestyle{empty}

\title{Constraints on axion-like particles and non-Newtonian gravity
from measuring the difference of Casimir forces}

\author{
G.~L.~Klimchitskaya}
\affiliation{Central Astronomical Observatory at Pulkovo of the
Russian Academy of Sciences, Saint Petersburg,
196140, Russia}
\affiliation{Institute of Physics, Nanotechnology and
Telecommunications, Peter the Great Saint Petersburg
Polytechnic University, Saint Petersburg, 195251, Russia}

\author{
V.~M.~Mostepanenko}
\affiliation{Central Astronomical Observatory at Pulkovo of the
Russian Academy of Sciences, Saint Petersburg,
196140, Russia}
\affiliation{Institute of Physics, Nanotechnology and
Telecommunications, Peter the Great Saint Petersburg
Polytechnic University, Saint Petersburg, 195251, Russia}
\affiliation{Kazan Federal University, Kazan, 420008, Russia}

\begin{abstract}
We derive constraints on the coupling constants of axion-like particles
to nucleons and on the Yukawa-type corrections to Newton's gravitational law
from the results of recent experiment on measuring the difference of Casimir
forces between a Ni-coated sphere and Au and Ni sectors of a structured disc.
Over the wide range of axion masses from 2.61\,meV to 0.9\,eV the obtained
constraints on the axion-to-nucleon coupling are up to a factor of 14.6
stronger than all previously known constraints following from experiments
on measuring the Casimir interaction. The constraints on non-Newtonian
gravity found here are also stronger than all that following from the Casimir and
Cavendish-type experiments over the interaction range from 30\,nm to $5.4\,\mu$m.
They are up to a factor of 177 stronger than the constraints derived recently
from measuring the difference of lateral forces. Our constraints
confirm previous somewhat stronger limits obtained from the isoelectronic
experiment, where the contribution of the Casimir force was nullified.
\end{abstract}

\maketitle

\section{Introduction}

Both the scalar and pseudoscalar particles are predicted in many extensions
of the standard model \cite{1}. The light pseudoscalar particles, axions,
and different kinds of axion-like particles play an important role by
explaining the absence of both large electric dipole moment of a neutron
and strong CP violation in QCD \cite{1a,2,3}. In astrophysics and cosmology
axions are considered as the most probable constituents of dark matter
\cite{4,4a,4b,5,5a}.
An exchange of light scalar particles between atoms of two closely spaced
macrobodies leads to the Yukawa-type corrections to Newton's gravitational law
\cite{6}. Similar corrections have been predicted by the extra-dimensional
unification schemes with a low-energy compactification scale \cite{7,8}.
It is important to remember that at separations below a few micrometers the
corrections of Yukawa-type, which far exceed the Newtonian gravitation, are
not excluded experimentally. Of special interest is the hypothetical scalar
particle called chameleon whose mass depends on the matter density of an
environment \cite{9,10}. This particle may be considered as a constituent of
dark energy and is discussed in connection with the observed late-time
acceleration of the Universe expansion \cite{11,12}.

In spite of many attempts, non of the predicted light pseudoscalar and scalar
particles has been discovered so far. Specifically, axion-like particles
have been searched in many laboratory experiments using their interactions with
photons, electrons and nucleons (see, e.g., reviews in Refs.~\cite{4,13,14,15}),
in astrophysical observations \cite{14,15,16,16a},  and in gravitational
experiments \cite{17,18}. The gravitational experiments of E\"{o}tvos and
Cavendish type were also used to constrain the Yukawa-type corrections to
Newtonian gravity mediated by the scalar particles (see Refs.~\cite{6,19} for
a review and one more recent experiment in Ref.~\cite{20}).

As was proposed long ago \cite{21,22}, measurements of the van der Waals and
Casimir forces can be used for constraining different corrections to Newton's
gravitational law. During the last few years these forces have been under an
active study both experimentally and theoretically (see Refs.~\cite{23,24} for
a review). The strongest constraints on the Yukawa-type corrections to Newtonian
gravity, following from the most precise measurements of the Casimir interaction,
have been obtained in Refs.~\cite{25,26,27,28,29} in the interaction range below
a micrometer. Recently, the major strengthening was achieved in the isoelectronic
experiment, where the Casimir force, acting perpendicular to the test surfaces,
was nullified \cite{30} (see also the version of isoelectronic experiment
based on measuring the difference of lateral forces \cite{31}).

Furthermore, it was shown that precise experiments on measuring the Casimir-Polder
and Casimir forces place strong limits on the coupling constants of axion-like
particles to nucleons \cite{32,33,34,35}. For the axion-like particles, which are
lighter than 1\,eV, even stronger constraints have been derived \cite{36} from
the isoelectronic experiment of Ref.~\cite{30}. According to the proposal of
Ref.~\cite{37}, the model-independent constraints on an axion are obtainable from
measuring the Casimir force between two test bodies with aligned nuclear spins.
It was shown also \cite{38,39} that the best laboratory constraints on the
parameters of a chameleon can be obtained from precise measurements of the Casimir
force. Experiments of this kind have been proposed in Refs.~\cite{40,41}.

In this paper, we derive the constraints on the coupling constants of axion-like
particles to nucleons and on the Yukawa-type corrections to Newtonian gravity
from the recent experiment on measuring the difference of Casimir forces \cite{42}
between a Ni-coated sphere and  Au and Ni sectors of the structured disc.
This disc consisted of alternating Ni and Au sectors deposited on a Si substrate.
It was covered by two sufficiently thin homogeneous overlayers made of Ti and Au.

The differential measurements of the Casimir force between metallic test bodies
in similar configurations have been proposed in Refs.~\cite{43,44,45} in order
to perform the conclusive test on the account of dissipation in the Lifshitz theory
of dispersion forces \cite{46,47}. The point is that the measurement data of several
precise experiments (see a review in Ref.~\cite{23} and more recent results
in Refs.~\cite{48,49,50}) have been found to exclude theoretical predictions of
the Lifshitz theory combined with the dielectric permittivity of the Drude model
taking into account the relaxation properties of free electrons. The same data
turned out to be in a very good agreement with theory if the dielectric permittivity
of the lossless plasma model is used at low frequencies. It should be noted,
however, that within the distance range of all precise experiments below a
micrometer the differences in theoretical predictions of both approaches do not
exceed a few percent. Because of this, the obtained results have been considered
by some authors as not enough convincing {see, e.g., Refs.~\cite{51,52} for
a discussion).

The situation has been changed recently after the experiment on measuring the
difference of Casimir forces \cite{42} was performed. In this experiment,
the alternative theoretical predictions using the Drude and the plasma models
differ by up to a factor of several thousands of percent. That is why an
unequivocal exclusion of the Lifshitz theory combined with the Drude model
and good agreement of the same theory using the plasma model, demonstrated
in the experiment \cite{42}, can be considered as conclusive.

Below, we use the differential measurement data of Ref.~\cite{42} to derive
the constraints on an axion and non-Newtonian gravity and compare them with
those following from the previously performed individual measurements of the
Casimir interaction. It is shown that over a wide interaction range the
obtained constraints are much stronger than all other constraints derived from
the Casimir experiments. The constraints on an axion, found here, are complementary
(up to a factor of 2 differences) to those of Ref.~\cite{30} following from
the isoelectronic experiment of Ref.~\cite{32}, where the Casimir force was
nullified. Our present constraints on the corrections to Newtonian gravity
are weaker by up to a factor of 10.5 than that derived in Ref.~\cite{30}, but
stronger by up to a factor of 177 than the constraints derived in recent
Ref.~\cite{31} exploiting measurements of the lateral force.

The paper is organized as follows. In Sec.~II we obtain constraints on the
coupling constants of axion-like particles to neutrons and protons from
measuring the difference of Casimir forces. In Sec.~III the same measurement
data are used to derive constraints on the Yukawa-type corrections to
Newton's gravitational law. Section~IV contains our conclusions and
discussion.

Throughout the paper we use units in which $\hbar=c=1$.

\section{Constraints on the coupling constants of axion-like particles
to nucleons}

We consider the axion-like particles interacting with nucleons (protons and
neutrons) via the pseudoscalar Lagrangian \cite{4}. Then, the effective
interaction potential between two nucleons situated at the points
$\mbox{\boldmath$r$}_1$ and  $\mbox{\boldmath$r$}_2$ belonging to two test
bodies (a sphere and a disc in our case)  arises due to the process of
two-axion exchange \cite{17,53,54}
\begin{equation}
V_{kl}(|\mbox{\boldmath$r$}_1-\mbox{\boldmath$r$}_2|)=
-\frac{g_{ak}^2g_{al}^2}{32\pi^3m^2}\,
\frac{m_a}{|\mbox{\boldmath$r$}_{12}|^2}\,
K_1(2m_a|\mbox{\boldmath$r$}_{12}|)
.
\label{eq1}
\end{equation}
\noindent
Here, $g_{ak}$ and $g_{al}$ are the axion-proton
($k,\,l=p$) or axion-neutron ($k,\,l=n$) dimensionless coupling constants,
 $m=(m_n+m_p)/2$ is the mean mass of a nucleon, $m_a$ is the mass of an axion,
$\mbox{\boldmath$r$}_{12}=\mbox{\boldmath$r$}_1-\mbox{\boldmath$r$}_2$,
 and $K_1(z)$ is the modified
Bessel function of the second kind. Note that Eq.~(\ref{eq1}) is
derived under the condition
$|\mbox{\boldmath$r$}_{12}|\gg 1/m$ which
is satisfied in all experiments on measuring the Casimir interaction.

In the experiment \cite{42} on measuring the difference of Casimir forces the
first test body was a sapphire (Al${}_2$O${}_3$) sphere (with a density
$\rho_s=4.1\times 10^3\,\mbox{kg/m}^3$) covered with the thermally evaporated
layers of Cr of thickness $\Delta_{\rm \,Cr}=10\,$nm
($\rho_{\rm Cr}=7.15\times 10^3\,\mbox{kg/m}^3$)
and Ni of thickness $\Delta_{\rm \,Ni}=250\,$nm
($\rho_{\rm Ni}=8.9\times 10^3\,\mbox{kg/m}^3$).
The Ni-covered sphere had a radius of $R=150.8\,\mu$m.
The second test body was the structured disc consisting of alternating Au and Ni
sectors. It was covered by the homogeneous Ti and Au
overlayers with thicknesses
$\Delta_{\rm \,Ti}=10\,$nm  and  $\Delta_{\rm \,Au}=21\,$nm, respectively.
These overlayers effectively enhance the variation in the difference of Casimir
forces between a Ni-coated sphere and sectors of the disc made of Au and Ni
when the Drude and plasma models are used in calculations. At the same time, the
homogeneous overlayers do not contribute to the difference of additional forces,
originating from either two-axion exchange or from the Yukawa-type corrections
to Newtonian gravity. To finish with a description of the second test body, we
note that the thickness of both Au
($\rho_{\rm Au}=19.31\times 10^3\,\mbox{kg/m}^3$)
and Ni sectors was $D=2.1\,\mu$m.
The structured disc covered with two overlayers was placed on the top of a
homogeneous Si wafer of thickness $\Delta_{\rm \,Si}=100\,\mu$m.
This wafer also does not contribute to the difference of additional forces.

The difference of additional forces between a sphere and a Au and a Ni sectors
of the structured disc due to two-axion exchange can be calculated using the
potential (\ref{eq1}). With appropriately replaced materials of the layers,
the result can be found in Ref.~\cite{36}
\begin{eqnarray}
&&
|\Delta F_{\rm diff}^{\rm add}(a)|=
\frac{\pi }{2m_am^2m_{\rm H}H^2}(C_{\rm Au}-C_{\rm Ni})
\label{eq2}\\
&&~~
\times\!\!
\int_{1}^{\infty}\!\!du\frac{\sqrt{u^2-1}}{u^3}e^{-2m_aua}\left(1-e^{-2m_auD}\right)
X(m_au),
\nonumber
\end{eqnarray}
\noindent
where $m_{\rm H}$ is the mass of an atomic hydrogen, $a$ is the distance between
a sphere and the sectors of a disc,  and the following notation is introduced
\begin{eqnarray}
&&
X(z)\equiv C_{\rm Ni}
\Phi(R,z)
\label{eq4} \\
&&
+(C_{\rm Cr}-C_{\rm Ni})
e^{-2z\Delta_{\rm Ni}}
\Phi(R-\Delta_{\rm Ni},z)
\nonumber \\
&&
+(C_{s}-C_{\rm Cr})e^{-2z(\Delta_{\rm Ni}+\Delta_{\rm Cr})}
\Phi(R-\Delta_{\rm Ni}-\Delta_{\rm Cr},z)
\nonumber
\end{eqnarray}
\noindent
with the function $\Phi$ defined as
\begin{equation}
\Phi(r,z)=r-\frac{1}{2z}+e^{-2rz}\left(r+
\frac{1}{2z}\right).
\label{eq4}
\end{equation}
\noindent

Here, the coefficients $C_{\rm M}$ with an index
 M=Au, Cr, $s$, and Ni,  for gold, chromium, sapphire and
 nickel, respectively,
are defined  as
\begin{equation}
C_{\rm M}=\rho_{\rm M}\left(\frac{g_{ap}^2}{4\pi}\,
\frac{Z_{\rm M}}{\mu_{\rm M}}+\frac{g_{an}^2}{4\pi}\,
\frac{N_{\rm M}}{\mu_{\rm M}}\right),
\label{eq5}
\end{equation}
\noindent
where $\rho_{\rm M}$ is the density, $Z_{\rm M}$ and
$N_{\rm M}$ are
the number of protons and the mean number of neutrons in an
atom or a molecule of the respective material, and
$\mu_{\rm M}=m_{\rm M}/m_H$,
$m_{\rm M}$  being the mean atomic (molecular) mass of the material M.
Note that the values of $Z/\mu$ and
$N/\mu$ for many elements
with account of their isotopic
composition are contained in Ref.~\cite{6}.
In our calculations below we use
$Z_{\rm M}/\mu_{\rm M}=0.40422$, 0.46518, 0.49422, and 0.48069 and
$N_{\rm M}/\mu_{\rm M}=0.60378$, 0.54379, 0.51412, and 0.52827 for
Au, Cr, sapphire, and Ni, respectively.

Now we obtain constraints on the coupling constants $g_{an}$ and
$g_{ap}$ from the experimental results of Ref.~\cite{42}.
For this purpose, we use the measurement set which was found in
agreement with theoretical results for the difference in
Casimir forces predicted by the Lifshitz theory and the plasma model
within the limits of $\Delta F=1\,$fN error over the separation range
from 250 to 400\,nm (see Fig.~12 of Ref.~\cite{42}). This means that
the difference of additional forces (\ref{eq2}) arising due to two-axion
exchange satisfies the condition
\begin{equation}
|\Delta F_{\rm diff}^{\rm add}(a)|<\Delta F.
\label{eq6}
\end{equation}

Note that the distances $a$ between the sphere and the sectors of a rotating
disc are connected with the experimental separations $z$ by
\begin{equation}
a=z+\Delta_{\rm\, Ti}+\Delta_{\rm\, Au}
=z+31\,\mbox{nm},
\label{eq7}
\end{equation}
\noindent
i.e., differ by the combined thickness of Ti and Au overlayers.
We have substituted Eqs.~(\ref{eq2})--(\ref{eq5}) in Eq.~(\ref{eq6})
and found numerically the values of $g_{an}$,
$g_{ap}$ and $m_a$ satisfying the inequality (\ref{eq6}) at different
separations $a$. In so doing, the most strong constraints have been obtained
at $a=291\,$nm ($z=260\,$nm).

In Fig.~\ref{fg1}, we present the computational results for allowed and
excluded regions of the plane $(m_a,g_{ap(n)}^2/4\pi)$ which lie below and
above each of the lines, respectively. The three lines from top to bottom are
plotted under the respective assumptions $g_{ap}^2\gg g_{an}^2$,
$g_{an}^2\gg g_{ap}^2$, and $g_{ap}^2= g_{an}^2$.

In Fig.~\ref{fg2}, we compare the constraints of Fig.~\ref{fg1} with the
strongest laboratory constraints on the coupling constants of axion-like
particles to nucleons obtained so far in the same region of axion masses.
The comparison is made under the plausible condition $g_{ap}= g_{an}$
\cite{17}. The line 1 shows the constraints obtained \cite{35} from
measurements of the lateral Casimir force between sinusoidally corrugated
surfaces \cite{55,56}. By the line 2 we present the constraints found \cite{34}
from measuring the effective Casimir pressure by means of micromechanical
torsional oscillator \cite{27,28}. The line 3 is obtained in this work using the
experiment \cite{42} on measuring the difference of Casimir forces.
It reproduces the bottom line in Fig.~\ref{fg1}. The constraints derived \cite{36}
from the isoelectronic experiment of Ref.~\cite{30}, where the Casimir force
was nullified, are shown by the line 4. Finally, the line 5 demonstrates the
constraints obtained \cite{18} from the Cavendish-type experiment of Ref.~\cite{57}.
The  regions of the plane $(m_a,g_{ap(n)}^2/4\pi)$ above each line are
experimentally excluded.

As is seen in Fig.~\ref{fg2}, the constraints of the line 3,
derived here from
measuring the difference of Casimir forces, are stronger than the gravitational
constraints and than all the other constraints obtained from measurements
of the Casimir force within the wide range of axion masses $m_a$ from 2.61\,meV
to 0.9\,eV. The maximum strengthening by a factor of 14.6 is achieved for
$m_a=4.88\,$meV. Up to a factor 2  are stronger constraints
given by the line 4 obtained from an experiment \cite{30},
where the Casimir force  was nullified. Thus, one can say that the constraints
of the line 3 confirm previous somewhat stronger constraints of
the line 4.

\section{Constraints on the Yukawa-type corrections to Newtonian gravity}

Now we consider two atoms with masses $m_1$ and $m_2$ situated at the points
$\mbox{\boldmath$r$}_1$ and  $\mbox{\boldmath$r$}_2$ of the test
bodies in the same experiment on measuring the difference of Casimir forces.
An exchange of one light scalar particle of mass $M=1/\lambda$ results in
the Yukawa-type effective potential, which is usually considered as a
correction to Newtonian gravitational potential  \cite{6}
\begin{equation}
V(|\mbox{\boldmath$r$}_{12}|)=
-\frac{Gm_1m_2}{|\mbox{\boldmath$r$}_{12}|}\,\left(1+\alpha
e^{-|\mbox{\boldmath$r$}_{12}|/\lambda}\right)
.
\label{eq8}
\end{equation}
\noindent
Here, $G$ is the Newtonian gravitational constant and $\alpha$ is a dimensionless constant
of the strength of Yukawa interaction. As mentioned in Sec.~I, the potential (\ref{eq8}) also
arises in extra-dimensional unification schemes with a low-energy compactification
scale \cite{7,8}.

A difference of the additional forces between a sphere and Au and Ni sectors of the
structured disc arising due to potential (\ref{eq8}) can be easily calculated as
described in Ref.~\cite{58}
\begin{eqnarray}
&&
|F_{\rm diff}^{\rm Yu}(a)|=4\pi^2G|\alpha|\lambda^3Re^{-{a}/{\lambda}}
(\rho_{\rm Au}-\rho_{\rm Ni})\left(1-e^{-{D}/{\lambda}}\right)
\nonumber \\
&&~~~~
\times\!\left[
\rho_{\rm Ni}
+(\rho_{\rm Cr}-\rho_{\rm Ni})e^{-\frac{\Delta_{\rm\, Ni}}{\lambda}}
\right.
\nonumber \\
&&~~~~~~~~~~~~\left.
+(\rho_{s}-\rho_{\rm Cr})e^{-\frac{\Delta_{\rm\, Ni}+\Delta_{\rm\, Cr}}{\lambda}}\right].
\label{eq9}
\end{eqnarray}
\noindent
The constraints on the parameters $\alpha$ and $\lambda$ of the potential (\ref{eq8})
are obtained from the measurement set of Ref.~\cite{42} specified in Sec.~II.
For this purpose, a difference of the Yukawa-type additional forces (\ref{eq9})
was substituted in Eq.~(\ref{eq6}) in place of $F_{\rm diff}^{\rm add}$ and the
numerical analysis of the obtained inequality has been performed.
The strongest constraints on the parameters $\alpha,\>\lambda$ were obtained
at the same separation distance $a=291\,$nm, as in Sec.~II.

In Fig.~\ref{fg3}, our constraints on the Yukawa-type corrections to Newtonian
gravity, following from measuring the difference of Casimir forces in Ref.~\cite{42},
are shown by the line 6. For comparison purposes, in Fig.~\ref{fg3}
the other strongest laboratory
constraints are shown in the same interaction range. The line 1 indicates the constraints
following from measurements of the effective Casimir pressure \cite{27,28}.
The constraints of the line 2 were obtained from the previous isoelectronic
(Casimir-less) experiment \cite{59}. The line 3 demonstrates the constraints
derived very recently from measuring the difference in lateral forces \cite{31}.
The line 4 shows the constraints obtained from measuring the Casimir force by means
of torsion pendulum \cite{29}. The constraints of the line 5 have been obtained from
the short-separation Cavendish-type experiment \cite{60,60a,61}. Finally, the line 7
represents the constraints derived from the results of recent isoelectronic
experiment of Ref.~\cite{30}. In all cases the regions of the plane
$(\lambda,|\alpha|)$ above each line are excluded by the results of respective
experiment, and the regions below each line are allowed.

As can be seen in Fig.~\ref{fg3}, the constraints of the line 6 obtained here from the
experiment \cite{42} on measuring the difference of Casimir forces are
quite competitive
over the wide interaction region from $\lambda=30\,$nm to $\lambda=5.4\,\mu$m.
In this region they are much stronger than all
the constraints obtained from other
measurements of the Casimir force. Specifically, the constraints of line 6 are up to
a factor of 16 stronger than that of line 1 derived from measurements of the
Casimir pressure. The maximum strengthening holds at $\lambda=80\,$nm.
The constraints of line 6 are also stronger than that following from the previous
isoelectronic (Casimir-less) experiment (line 2) and recent experiment on measuring the
difference of lateral forces (line 3). The maximum strengthenings by the factors of 122 and
177 hold at $\lambda=435\,$nm and $2\,\mu$m, respectively.
At $\lambda=3.1\,\mu$m the constraints of line 6 are stronger by the
factor of 100
than that of lines 4 and 5.
Our constraints turn out to be stronger than the ones
obtained from the Cavendish-type experiment (line 5) in the
interaction region $\lambda<5.4\,\mu$m.

At $\lambda=30\,$nm the obtained here constraints are of the same strength as those
found from the improved isoelectronic experiment of Ref.~\cite{30}. At larger
$\lambda$ the latter becomes stronger than the constraints of line 6 by up to a factor
of 10.5. This is explained by the fact that the force sensitivity in the experiment
\cite{30} is up to an order of magnitude higher than the measure of agreement
between the experimental force differences and theoretical predictions in Ref.~\cite{42}.
Therefore, both the isoelectronic experiment \cite{30} and the experiment on measuring
the difference of Casimir forces can be considered as two independent confirmations for
the obtained stronger constraints on the Yukawa-type corrections to Newtonian gravity.

\section{Conclusions and discussion}

In the foregoing, we have derived the constraints on the coupling constants
of axion-like particles to nucleons and on the non-Newtonian gravity of Yukawa
type from the results of recent experiment on measuring
the difference of Casimir forces \cite{42}. This experiment occupies a highly
important place among numerous experiments on measuring the Casimir interaction
because the predicted force difference vary by thousands of percent depending on the
used model of dissipation of free electrons. An important feature of the employed
differential measurement scheme is also that the role of possible background
effects, such as the electrostatic patches, surface roughness, and variation of
optical properties of material boundaries is largely suppressed \cite{42}.
All this allowed a unequivocal exclusion for theoretical predictions of the
Lifshitz theory using the Drude model and a conclusive demonstration of an
agreement of the same theory using the plasma model with the experimental data.
The measure of this agreement was used here to obtain more strong constraints on the
axions and non-Newtonian gravity than those obtained from all previous measurements
of the Caimir force.

According to our results, the derived constraints on the coupling constant of
axion to nucleons over a wide range of axion masses from 2.61\,meV to 0.9\,eV
are stronger by up to a factor of 16 than all previously known constraints
following from the Casimir experiments. They confirm by a factor of 2 stronger
constraints obtained previously
from the isoelectronic experiment \cite{30} where the
Casimir force was nullified.

The constraints on the Yukawa-type corrections to Newton's law of gravitation,
derived here from the same experiment, are stronger than all that found from
the previously performed measurements of Casimir and gravitational interactions
in the range from 30\,nm to $5.4\,\mu$m. The achieved strengthening is by up to the
factors of 16 and 122 as compared to the experiment on measuring the effective
Casimir pressure \cite{27,28} and previous Casimir-less isoelectronic
measurement \cite{59}, respectively. Our present constraints are by up to the factor
of 177 stronger than the results obtained very recently from measuring the difference
of lateral forces \cite{31}, but by up to a factor
10.5 weaker than the constraints
following from the latest version of the isoelectronic experiment \cite{30}.
Thus, at the moment the experiment on measuring the difference of Casimir
forces and the isoelectronic experiment lead to the
strongest constraints on both
the coupling constants of axions to nucleons and on
the Yukawa-type corrections
to Newtonian gravity over the respective regions of axion masses
$m_a$
and interaction lengths $\lambda$ indicated above.

\section*{Acknowledgments}

The authors are grateful to R.~S.~Decca for providing the numerical data
of line 7 in Fig.~\ref{fg3}.
The work of V.M.M. was partially supported by the Russian Government
Program of Competitive Growth of Kazan Federal University.

\newpage
\widetext
\begin{figure}[b]
\vspace*{-20cm}
\centerline{\hspace*{2.5cm}
\includegraphics{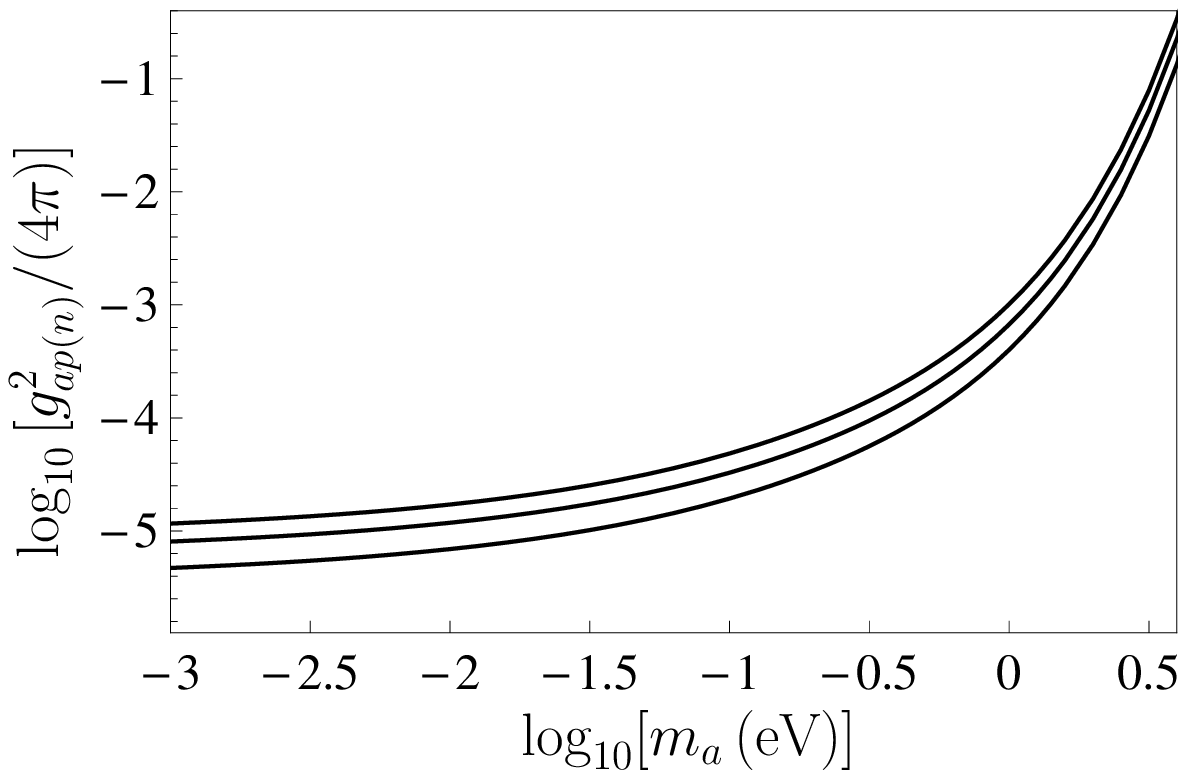}
}
\vspace*{-9cm}
\caption{\label{fg1}
The lines from top to bottom show the constraints on the coupling
constants of axion-like particles to a proton and a neutron as
functions of the axion mass, which follow from measuring the
difference of Casimir forces under the assumptions
$g_{ap}^2\gg g_{an}^2$,
$g_{an}^2\gg g_{ap}^2$, and $g_{ap}^2= g_{an}^2$, respectively.
The regions of the plane above each line are excluded and below
each line are allowed.
}
\end{figure}
\begin{figure}[b]
\vspace*{-8cm}
\centerline{\hspace*{2.5cm}
\includegraphics{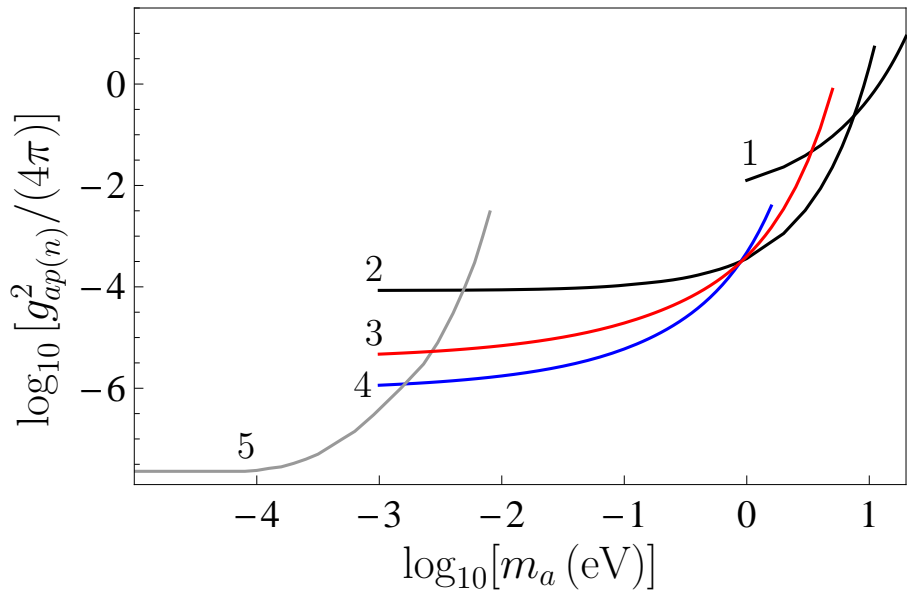}
}
\vspace*{-9cm}
\caption{\label{fg2}
The line 3 shows the constraints on the coupling
constants of axion-like particles to a proton and a neutron as
a function of the axion mass obtained here from measurements of
 the difference of Casimir forces.
The other lines show previous constraints derived
from measuring the
lateral Casimir force (line 1), the effective Casimir pressure
(line 2), from the isoelectronic
and Cavendish-type experiments (lines 4 and 5).
See the text for further discussion.
The regions of the plane above each line are excluded and below
each line are allowed.
}
\end{figure}
\begin{figure}[b]
\vspace*{-7cm}
\centerline{\hspace*{2.5cm}
\includegraphics{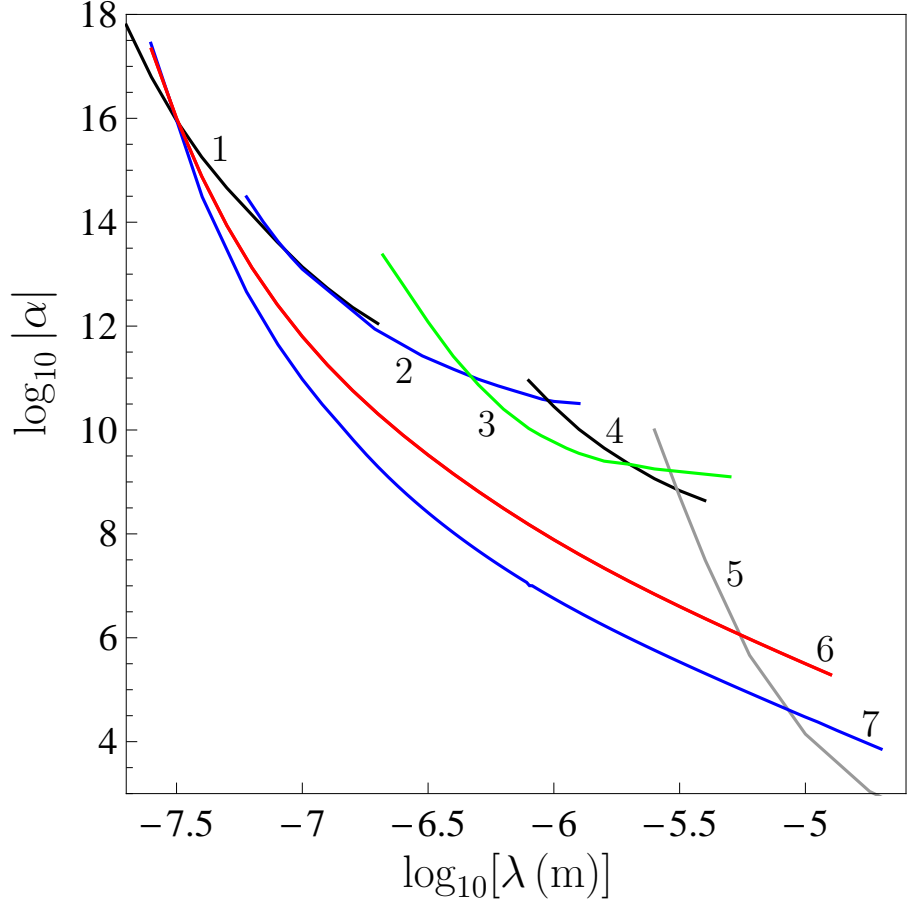}
}
\vspace*{-9cm}
\caption{\label{fg3}
The line 6 shows the constraints on the strength of
Yukawa-type correction to Newton's gravitational law as a
function of the interaction length obtained here
from the experiment on measuring the difference of Casimir forces.
The other lines show previous constraints derived from
measuring
 the effective Casimir pressure (line 1), from  previous isoelectronic
(Casimir-less) experiment (line 2), from experiment on measuring the
difference of lateral forces (line 3), from measuring the Casimir force
by means of torsion pendulum (line 4),
from the Cavendish-type experiments
(line 5),
 and from the recent  isoelectronic experiment (line 7).
The regions of the plane above each line are excluded and below
each line are allowed.
}
\end{figure}
\end{document}